\documentclass[reprint,showpacs,superscriptaddress]{revtex4-1}

\pdfoutput=1

\usepackage{amsmath}
\usepackage{amsfonts}
\usepackage{amssymb}
\usepackage{graphicx}
\usepackage{braket}
\usepackage[colorlinks=true, allcolors=blue]{hyperref}

\begin{document}
\title{Dispersive optomechanics of supercavity modes in high-index disks}
\author{Laura Mercad\'e}
\email{laumermo@ntc.upv.es}
\affiliation{Nanophotonics Technology Center, Universitat Polit\`ecnica de Val\`encia, Camino de Vera s/n, 46022 Valencia, Spain}
\author{\'Angela Barreda}
\affiliation{Nanophotonics Technology Center, Universitat Polit\`ecnica de Val\`encia, Camino de Vera s/n, 46022 Valencia, Spain}
\affiliation{Institute of Applied Physics, Abbe Center of Photonics, Friedrich Schiller University Jena, Albert-Einstein-Str. 15, 07745 Jena, Germany}
\author{Alejandro Mart\'inez}
\affiliation{Nanophotonics Technology Center, Universitat Polit\`ecnica de Val\`encia, Camino de Vera s/n, 46022 Valencia, Spain}
\date{\today}
\begin{abstract}
In this work, we study the dispersive coupling between optical quasi-bound-states in the continuum at telecom wavelengths and GHz-mechanical modes in high-index wavelength-sized disks. We show that such cavities can display values of the optomechanical coupling rate on par with optomechanical crystal cavities ($g_{0}/2\pi\backsimeq$ 800 kHz). Interestingly, optomechanical coupling of optical resonances with mechanical modes at frequencies well above 10 GHz seems attainable. We also show that mechanical leakage in the substrate can be extremely reduced by placing the disk over a thin silica pedestal. Our results suggest a new route for ultra compact optomechanical cavities which can potentially be arranged in massive arrays forming optomechanical metasurfaces. 
\end{abstract}

\maketitle
Interaction between optical and mechanical waves in deformable cavities lies at the heart of cavity optomechanics \cite{ASP14-RMP}. Amongst the different approaches to build optomechanical (OM) cavities, their realization on high refractive index films using standard lithographic processes has received special interest. By introducing periodicity in the high-index films, it is possible to create OM bandgaps that forbid propagation of optical and mechanical waves at certain frequencies, resulting in the so-called optomechanical crystals (OMCs) \cite{EIC09-NAT}. Then, a wavelength-sized defect can allocate both optical and mechanical modes in a same spatial region ensuring strong overlap between localized fields \cite{PEN14-NP}. Indeed, this approach is the one that allows to confine both kinds of waves in a smaller volume (roughly, around $(\lambda/2n)^{3}$, where $\lambda$ is the free-space optical wavelength and $n$ is the refractive index of the confining material). This results in large values of the OM coupling rate $g_{0}/2\pi$ - the parameter that accounts for the OM interaction in the cavity – which can reach values higher than 1 MHz \cite{MATH18-APL}. The achievable Q factors are not so high as in the case of free-space Fabry Perot cavities \cite{THO08-NAT} or silica toroids \cite{VER12-NAT} but still large enough ($\approx 10^{5}$) as to operate in the so-called single sideband regime that enables the observation of intriguing phenomena such as ground-state cooling \cite{CHAN11-NAT}. Finally, this technological approach permits the on-chip integration of the OMCs with electro-mechanical devices \cite{PIT20-ARX}, which ultimately enable bidirectional and coherent microwave-optics transducers \cite{VAIN16-APL}. 

In parallel, the field of high-index nanophotonics has come up with new strategies to build wavelength-size optical cavities in high-index films without using periodic mirrors. By properly tailoring the aspect ratio of a high-index disk, interference between different optical modes cancels out the far-field scattering and enables tight localization of the optical field inside the disk. Moreover, the so-called quasi-bound states in the continuum (quasi-BICs) can arise, resulting in relatively high values of the Q factor (Q$>$100) \cite{RYB17-PRL,KOS20-SCI}. Notably, the strong field localization in single disks extremely enhances nonlinear effects such as second \cite{KOS20-SCI} or third harmonic generation \cite{GRI16-NL}. However, another nonlinear process such as OM interaction has not been studied so far in such structures. 

Here, we study the dispersive OM coupling in high-index disks supporting quasi-BIC modes at telecom wavelengths. We show that the disks also support mechanical resonances at microwave frequencies and they can be coupled to the optical resonance with significantly high values of $g_{0}$. We also show that having the disk supported by a thin silica pedestal, which is technologically feasible, would strongly suppress phonon leakage in the substrate ensuring material-limited mechanical losses, as in released OM cavities operating at room temperature. Our results suggest that these cavities could be interesting in applications where unresolved sideband operation is not mandatory as well as to build massive arrays of OM cavities to create GHz-tunable OM metasurfaces for multiple applications.

\begin{figure}[htbp]
\centering
\includegraphics[width=\linewidth]{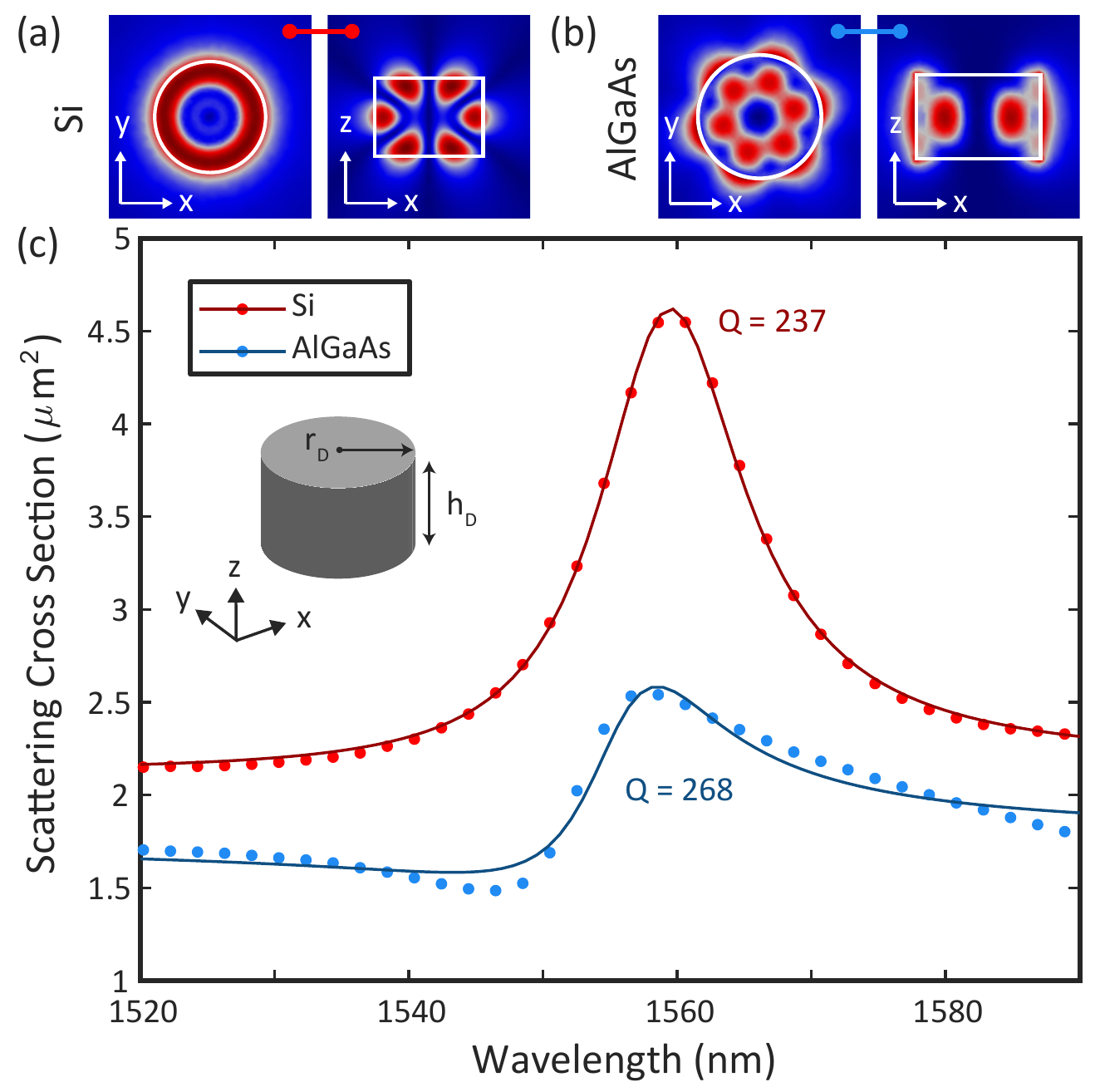}
\caption{Supercavity or quasi-BIC mode profiles of the total electric field $\vert \textbf{E}\vert^{2}$ for (a) a Si disk with h$_{d}$=589 nm and r$_{d}$=403 nm  at $\lambda$= 1559.1 nm and (b) an AlGaAs disk with h$_{d}$=635 nm and r$_{d}$=451.5 nm at $\lambda$= 1555.8 nm. (c) Simulated far field scattering cross section and Fano fits for the calculation of the estimated optical quality factor. Data points correspond to the simulated far-field scattering and the solid lines to the resulting fitted Fano resonance.}
\label{fig:scattering}
\end{figure}

Supercavity or quasi-BIC modes arise in wavelength-sized high-index dielectric disks as a result of the interference between different modes that allows to reduce radiation therefore achieving large Q factors. In Ref \cite{RYB17-PRL} it is predicted that a silicon disk with aspect radio $h/d$ = 0.7308, being h the disk height and d the disk diameter, can reach Q $\approx$ 200 at telecom wavelengths due to the interference between a Mie--type and a Fabry--Perot--type modes. The electric field patterns at $\lambda$= 1559.1 nm of this quasi--BIC mode are shown in Fig. \ref{fig:scattering}(a). The computed far-field scattering, shown on Fig. \ref{fig:scattering}(c), results in Q $\backsimeq$ 237. We also computed the field patterns at $\lambda$= 1555.8 nm of the quasi--BIC mode observed in Ref. \cite{KOS20-SCI} for a AlGaAs disk, assuming that it is completely surrounded by air. The results are depicted on Fig. \ref{fig:scattering}(b). Here, the quasi--BIC mode results from the interference between Mie--like modes, and the scattering (Fig. \ref{fig:scattering}(c)) shows an optical quality factor of Q $\backsimeq$ 268. The quality factor were estimated through a fit of the obtained scattering cross-section to a Fano resonance.

\begin{figure}[htbp]
\centering
\includegraphics[width=\linewidth]{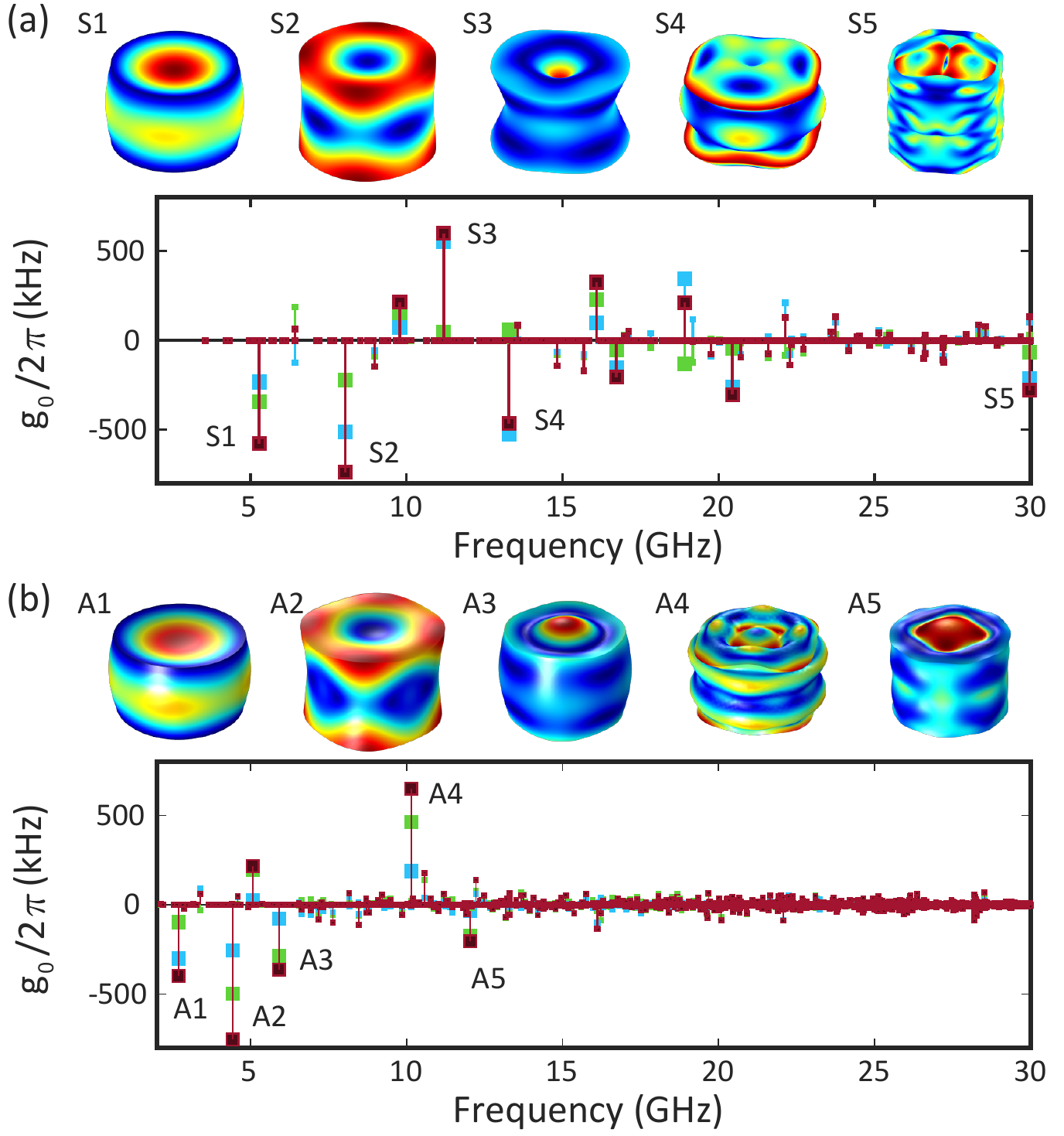}
\caption{OM coupling rate ($g_{0}/2\pi$) and mechanical mode profile (blue minimum and red maximum) for a (a) Si disk with h$_{d}$=589 nm and r$_{d}$=403 nm and (b) AlGaAs disk with h$_{d}$=635 nm and r$_{d}$=451.5 nm . The total coupling rate, and the photoelastic and moving boundary contributions are depicted in red, green and blue respectively. }
\label{fig:g0}
\end{figure}

We also computed the mechanical resonances of both disks up to 30 GHz using COMSOL Multiphysics. Once the displacement field was obtained, we were able to calculate the OM coupling rate $g_{0}$ for each mechanical mode from both the moving boundary (MB) and the photoelastic (PE) effect contributions, presented in Fig. \ref{fig:g0} in blue and green respectively. In our calculations, we assumed that our cavities can be well approximated as a closed system because of the relatively high optical Q. In other words, we neglected dissipative coupling and only accounted for dispersive coupling using the analytical expressions in Ref. \cite{CHAN12-APL}. Both the Si and AlGaAs materials were simulated as anisotropic media whose properties are summarized in Table \ref{tab:mec_properties} \cite{BAK14-OE, ADA85-JAP}. Regarding the photoelastic constants we used $p_{11}$=-0.09, $p_{12}$=0.017 and $p_{44}=-0.051$ for Si and $p_{11}$=-0.165, $p_{12}$=-0.140 and $p_{44}=-0.072$ \cite{BAK14-OE} for AlGaAs . Notice that the photoelastic coefficients of the AlGaAs are assumed to be equal to those of GaAs as they have been unknown in the literature. 

\begin{table}[htbp]
\centering
\caption{\bf Optical and mechanical properties of the employed materials}
\begin{tabular}{cccccc}
\hline
 & $\varepsilon $ & $\rho$   & $c_{11}$  & $c_{12}$ & $c_{44}$ \\
  &  &  (kg/m$^{2}$)  & (GPa) & (GPa) &  (GPa)\\
\hline
Si & 13 & 2330 & 166  & 64 & 80\\
AlGaAs & 11.42 & 4540 & 119.5 & 55.4 & 59.1\\

\hline
\end{tabular}
  \label{tab:mec_properties}
\end{table}

Figure 2(a) and (b) show the obtained values of $g_{0}/2\pi$ for the Si and AlGaAs disks, respectively, as well as the mechanical displacement pattern of the mechanical modes showing coupling rates higher than 200 kHz. Remarkably, $g_{0}/2\pi \approx$ 800 KHz, similar to what is obtained in OMCs implemented on 1D periodic beams, is obtained for different mechanical modes. It can be noted that in both cases the highest OM coupling rates with values of -737 kHz and -754 kHz for Si and AlGaAs, respectively, are obtained for the same mechanical mode profile depicted in S2 and A2. The mechanical frecuency of this mode varies from one disk to the other, but higher frequency modes can be found in the Si disk. Remarkably, a mechanical mode at 29.97 GHz with $g_{0}/2\pi$= 279 kHz can even been found in the Si disk (mode S5). 

\begin{figure}[htbp]
\centering
\includegraphics[width=\linewidth]{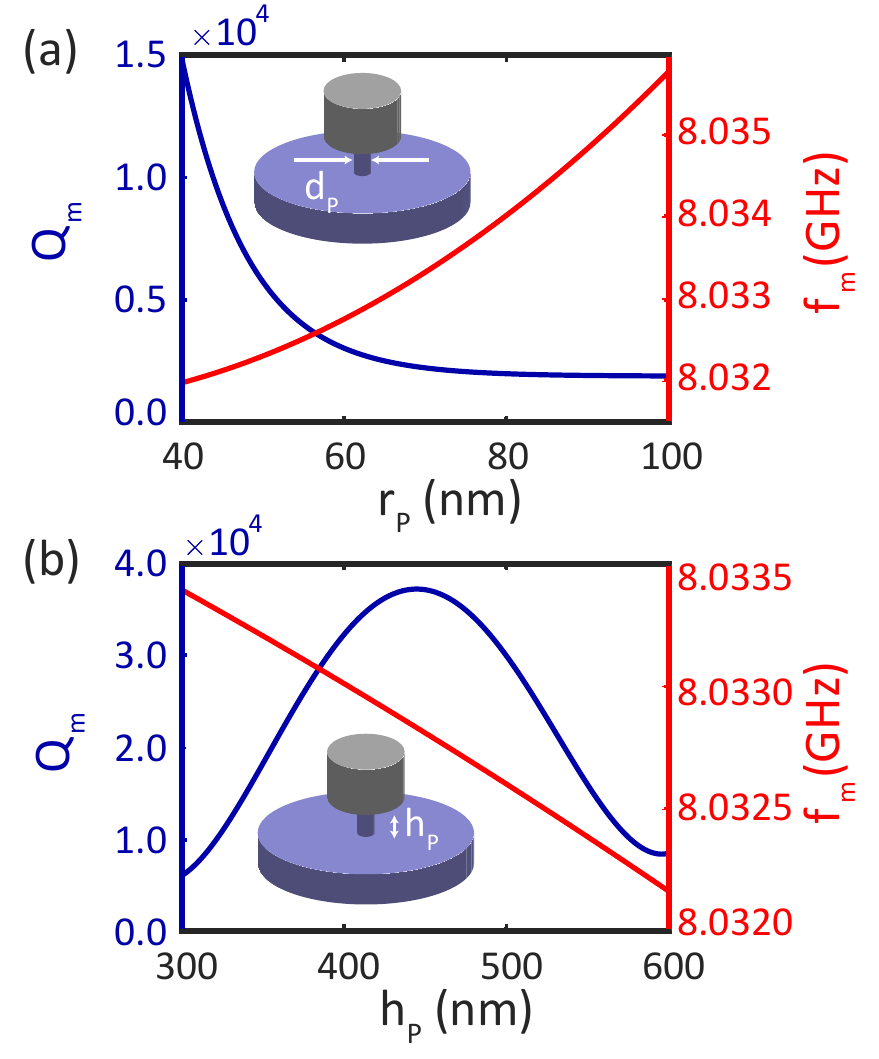}
\caption{Mechanical quality factor evolution and frequency shift of S2 as a function of the pedestal (a) radius with h$_{p}$= 500 nm and (b) height with d$_{p}$/2=r$_{p}$= 50 nm.}
\label{fig:meclosses}
\end{figure}

In the previous simulations, we have assumed that the disks are completely surrounded by air. This ensures that the mechanical vibrations are completely bounded to the disk and the phonon lifetime will be uniquely limited by the absorption losses in the material. However, this approach is not realistic, since in a real implementation the high-index disks must be placed on a substrate, typically silicon dioxide (SiO$_{2}$). Since SiO$_{2}$ is a low-index material, it can be expected that the presence of the substrate will have low influence on the optical properties. Moreover, the Q factor can be even improved by a further engineering of the surroundings of the disks \cite{RYB17-PRL}. However, sound velocity in SiO$_{2}$ is smaller than in either Si (or AlGaAs), which means that phonons will tend to leak towards the SiO2 substrate, resulting in prohibitive mechanical losses. This is the reason why released structures are conventionally used in integrated OMCs. Phonon leakage can be highly reduced if we assume the disk to be placed on a thin pedestal, following the approach presented in Ref. \cite{GIL15-NN} for GaAs disks and in Ref. \cite{LAER15-NP} for high-index waveguides. Thus, we assumed that the high-index disks were staying on a thin SiO$_{2}$ cylindrical pedestal (height $h_{p}$ and diameter $d_{p}$ as depicted in the insets of in Fig. \ref{fig:meclosses}). We studied the evolution of the mechanical quality factor Q$_{m}$as a function of the pedestal height and diameter for the S2 mode in the Si disk, the one providing the highest OM coupling rate. We neglected material absorption losses being our objective to estimate the structural mechanical Q factor. It has to be noted that in this configuration only the mechanical modes with a negligible displacement in the middle of the disk will still be localized. In fact, the localization factor defined as the total displacement ($\textbf{s}$) localized in the disk (I) divided by total displacement in all the structure (high-index disk and SiO$_{2}$ pedestal and substrate) (II) $\int_{I}\vert \textbf{s}^{2}\vert d\textbf{s}/\int_{II}\vert \textbf{s}^{2}\vert d\textbf{s}$ for S2 results approximately in 0.97 (h$_{p}$=500 nm and r$_{p}$=50 nm), which means that the mechanical mode is almost totally confined (a total confined mode should give a localization of 1) in the disk even when it is placed at the top of the pedestal. Figure \ref{fig:meclosses} summarizes the evolution of the Q$_{m}$ as well as the shift in frequency of the mechanical mode as a function of the height and radius of the SiO$_{2}$ pedestal. To calculate the mechanical quality factor mechanical Perfectly Matched Layers (PMLs) were imposed in the bottom and the surroundings of the substrate. The interfaces of the disk, the pedestal and the top of the substrate stayed as free boundary conditions. 

As shown in Fig. \ref{fig:meclosses}(a) an increase of pedestal radius results in a reduction of the mechanical Q factor, as expected. This is due to the fact that the S2 mode profile will be disturbed is the pedestal diameter extends beyonds the zero displacement region localized at the middle of the disk. In that case, the mechanical mode could leak into the substrate through the pedestal thus reducing the mechanical quality factor. If we assume that Q$_{m}\simeq$ 10$^3$ at room temperature due to material absorption \cite{GOMIS14-NCOMM}, this will be the dominant mechanism for phonon losses for very thin radii. In other words, assuming r$_{p}$ < 80 nm and h$_{p}$ $\backsimeq$ 450 nm the mechanical properties will be the same as in the case of released OMC cavities when operating at room temperature. In Fig. \ref{fig:meclosses}(b) we can see that there is an optimum value of h$_{p}$ that maximices Q$_{m}$. We believe that this is due to a resonant effect in the pedestal that contributes to maximize the mechanical field in the disk when h$_{p}$ = 450 nm.

The previous results suggest that supercavity modes in high-index disk may play an important role in cavity optomechanics. Even though operation in the sideband-resolved regime seems unattainable, the large values of $g_{0}$ would ensure transduction of the GHz-scale mechanical modes as well as its manipulation either by cooling or heating them. In particular, blue-detuned laser driving at high power could ensure the formation of broad ($>$ 100 GHz) optical frequency combs \cite{MER19-ARX} in a wavelength-sized structure, opening the door towards ultra-compact microwave photonics. Notice also than the dimensions of the disks under study are smaller than those reported in \cite{GIL15-NN,BAK14-OE}, which support high-Q optical and mechanical whispering gallery modes. This further miniaturization to sizes of the order of the wavelength allow for accessing mechanical modes at higher frequencies, going even beyond 10 GHz, which is highly challenging in OMC cavities \cite{SAF14-PRL}.

In our study, we have only considered dispersive coupling even though the system presents nonnegligible radiation losses. Still, we have observed small variations ($<$ 1 $kHz$) of the calculated $g_{0}$ when the lateral size of the simulation volume is changed by a factor of 10 times higher. This means that our results should be considered as a good approximation to the dispersive behaviour of the system. Further calculations using quasi-normal modes \cite{WEI20-PRL} would enable to calculate the dissipative coupling, which could shed more light on the OM properties of this kind of systems. 

In comparison with OMC cavities, high-index disk can be excited from the far-field – though by using special arrangements of polarization \cite{KOS20-SCI}, which eases the driving conditions. This way, we may think on OM metasurfaces formed by arrays of such disk so that their properties could be dynamically tuned at sub-nanosecond speeds, leading to ultrafast reconfigurable metasurfaces. The high mechanical frequencies would also ensure operation in liquid environments relevant in biology and chemistry \cite{GIL15-NN}. This could enable to detect organisms smaller than bacteria, such as viruses or proteins, in on-chip optomechanical arrays \cite{GIL20-NN}.

\section*{Funding Information}


This work was supported by the European Commission (PHENOMEN H2020-EU-713450) and THOR H2020-EU-829067; Programa de Ayudas de Investigacin y Desarrolo (PAID-01-16)
de la Universitat Polit\`ecnica de Val\`encia; Ministerio de Ciencia, Innovacion y Universidades (PGC2018-
094490-B-C22) and Generalitat Valenciana (PROMETEO/2019/123, PPC /2018 /002, IDIFEDER/2018/033) and BEST/2020/178.
A.B. acknowledges the funding support from the Humboldt Research Fellowship from the Alexander von Humboldt Foundation.

\bibliography{biblio.bib}

\end{document}